\documentclass[%
 reprint,
 amsmath,amssymb,
 aps,
prb,
]{revtex4-2}

\usepackage{graphicx}
\usepackage{dcolumn}
\usepackage{bm}

\begin{document}

\preprint{APS/123-QED}

\title{Superconductivity in Ca-intercalated bilayer silicene}

\author{Jisvin Sam$^1$, Sasmita Mohakud$^2$, Katsunori Wakabayashi$^3$ and Sudipta Dutta$^1$}
\affiliation{$^1$Department of Physics, Indian Institute of Science Education and Research (IISER) Tirupati, Tirupati - 517507, Andhra Pradesh, India \\
$^2$Department of Physics, School of Advanced Sciences, Vellore Institute of Technology, Vellore, Tamil Nadu, 632014, India \\
$^3$Department of Nanotechnology for Sustainable Energy, School of Science and Technology, Kwansei Gakuin University, Sanda, Hyogo 669-1330, Japan
}

\date{\today}

\begin{abstract}
Within first-principles calculations, we explore superconductivity in Ca-intercalated bilayer silicene compound, Si$_{2}$CaSi$_{2}$. This arises from the coupling of interlayer flower-like $\rm \Gamma$-centered Fermi surface formed by the hybridization of Ca-3d and Si-3p$_{z}$ orbitals with low-energy out-of-plane vibrations enabled by silicene's buckling. The consequent large electron-phonon coupling, as evident from the Eliashberg spectral function leads to superconductivity below 5.4 K in this two-dimensional covalent system. Our results reveal the key control parameters to achieve superconductivity in experimentally synthesizable silicon-based thin materials that can find diverse applications.

\end{abstract}

\maketitle



Superconductivity in atomically thin materials are of sustained interest both from fundamental and application perspectives. Such superconductors yield notable enhancements in critical properties, including the critical magnetic field and current, when the coherence length becomes comparable with the system size\cite{nr1,nr2}. Recent experimental advancements have made possible the realization of a handful of atomically thin two-dimensional (2D) materials with mostly semi-metallic and semiconducting behaviors. There have been continued efforts to realize the long-range order properties like superconductivity in such systems by various means\cite{nr3}. 

Intercalation of alkali or alkaline earth metal in bulk graphite, known as graphitic intercalated compounds (GICs) have been known to be superconducting since the early 1960s \cite{nr4}. Ca-intercalated graphite with C$_{6}$Ca stoichiometry has been reported to have the highest superconducting transition temperature, $T_{c}$ = 11.5K among the GICs \cite{nr5}. Subsequently, Ca-intercalated bilayer graphene, i.e., C$_{6}$CaC$_{6}$ was reported to be superconducting with $T_{c}$ = 4K \cite{nr6}. The Li-decorated monolayer graphene has also been predicted \cite{nr7} to exhibit superconducting behavior. The superconductivity in the reported GICs is explained in terms of the occupancy of the intercalant-derived interlayer band and subsequent scattering of the electrons between the intercalant and the $\pi$-bands\cite{nr8,nr9}. The free electron-like parabolic interlayer band allows the electrons to couple with out-of-plane phonon vibrations, thereby increasing the electron-phonon coupling strength ($\lambda$). Based on these observations, one would expect the enhancement of the $\lambda$ in terms of increased out-of-plane phonon vibrations, if the flat graphene layer is replaced by a buckled honeycomb layer of similar band dispersions. This can result in enhancement of the $T_{c}$, that is desired for practical applications. 

This has motivated us to use the silicene atomic sheet in place of graphene\cite{nr9a}. The honeycomb lattice of silicene shows almost similar band dispersions like graphene near the Fermi energy, with slight gap opening\cite{nr9b}. However, unlike graphene, the silicene atomic sheet is a buckled structure with out-of-plane distortion of two Si atoms in the rhombus unit cell, that can enhance the $\lambda$. Due to the lack of mirror reflection symmetry in buckled silicene, the $\pi$ electrons can be scattered by low energy out-of-plane flexural modes, i.e., the ZA phonons\cite{nr10}. Large electron-phonon matrix elements for such scattering with low energy phonons increase the $\lambda$, favoring a large $T_{c}$. Mirror symmetry in graphene prohibits such scattering processes with $\pi$ electrons \cite{nr11}. Moreover, the superconductivity cannot be explained solely by the coupling of the $\pi$ electrons with the in-plane vibrations of the C atoms, i.e., the optical modes\cite{nr12}. Therefore, only the charge transfer by the intercalants and consequent increase in the density of states (DOS) near the Fermi energy are insufficient to induce superconductivity in graphene. Intercalated silicene can serve as a better alternative from that perspective.

Electron-doped silicene shows superconductivity under biaxial strain and the $T_{c}$ can be enhanced up to 10K with increase in applied strain, owing to an increase in the DOS at the Fermi level and the creation of new channels for electron-phonon coupling\cite{nr13}. Lattice mismatch induced effective strain on silicene from the Ag substrate, in addition to the charge transfer from Ag has been shown to exhibit superconducting-like charge gap\cite{nr14,nr15}. Charge transfer from the alkaline earth metals, e.g., Ca to silicene in case of rhombohedral CaSi$_2$ leads to superconductivity at 3K only under applied pressure\cite{nr16}. Under enhanced pressure, the system undergoes a phase transition to a flat configuration with an increase in $T_{c}$ to 14K, owing to a softening of the $B_{2g}$ optical phonon mode and increased $\pi^{*}$ electron density at Fermi energy\cite{nr17}. The electron-phonon coupling in bulk CaSi$_2$ is significantly influenced by the buckling of silicene and can be adjusted through external pressure or through the introduction of intercalant atoms.

We investigate the superconducting property of Ca-intercalated bilayer silicene, i.e., Si$_2$CaSi$_2$ in this paper. This system shows superconductivity at 5.4K in absence of any external pressure, owing to the occupied Ca-Si hybridized band and charge-transfer induced softening of the acoustic phonon modes.

 \begin{figure}[h]
\includegraphics[scale=0.4]{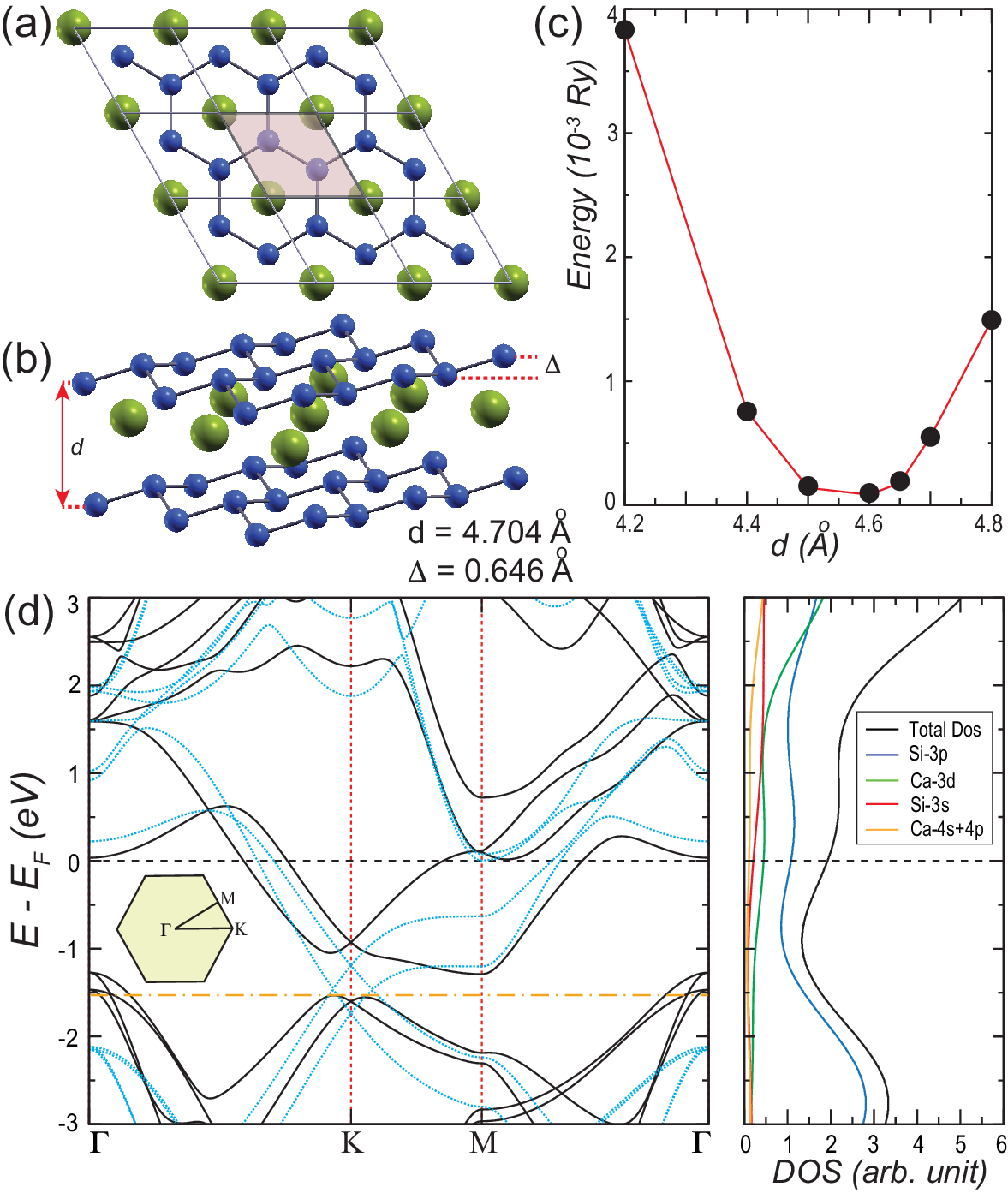}
\caption{\label{fig:epsart}  (color online) (a) The top view and (b) the side view of the Ca intercalated bilayer silicene system, Si$_2$CaSi$_2$. The unit cell is shown by shaded rhombus. The buckling of monolayer silicene and the vertical distance between two AA stacked silicene layers are denoted by $\rm \Delta$ and $d$, respectively. (c) The variation of the relative total energy of the system as a function of $d$, placing the Ca atom exactly at the middle of two silicene layers. (d) The band structure (solid lines) and DOS of the Si$_2$CaSi$_2$ system along with the orbital projected DOS, scaled with respect to the Fermi energy. The vertical dashed lines show the location of the high-symmetric points in the hexagonal Brillouin zone (inset). The horizontal dashed line depicts the Fermi energy of the system. The band structure of the bilayer silicene by removing the Ca atom from the sandwich compound is presented by dotted lines for comparison. Note that, this band structure is shifted by a constant energy, obtained by subtracting its Fermi energy (horizontal dashed-dotted line) from the Fermi energy of the Si$_2$CaSi$_2$ system.}
\end{figure}

We adopt ab-initio calculations using plane wave basis as implemented in Quantum Espresso\cite{nr18} with local density approximation (LDA)\cite{nr19}, using non-relativistic norm-conserving pseudopotential\cite{nr20}. A plane wave kinetic energy cutoff of 150 Ry and Methfessel-Paxton smearing of 0.09 Ry are considered along with a $\Gamma$-centered Brillouin 
zone (BZ) with $k$-point mesh of 30$\times$30$\times$1 for the electronic charge density calculations. The rhombus unit-cells of monolayer silicene and Si$_2$-Ca-Si$_2$ are relaxed till the total force converges at 3.4$\times$10$^{-5}$Ry/au. 

The phonon properties are calculated using density functional perturbation Theory (DFPT) with a 6$\times$6$\times$1 $q$-mesh. The electron-phonon coupling strength, $\lambda_{\rm \textbf{q}\nu}$ associated with a specific phonon mode $\nu$ and wave vector $\rm \textbf{q}$ is obtained from electron-phonon matrix element $g_{mn,\nu}(\rm \textbf{k},\rm \textbf{q})$, where $\rm \textbf{k},\rm \textbf{k+q}$ are the electron wave vectors of scattering state and $m$,$n$ are the electron band indices, as follows:

\begin{eqnarray}
\lambda_{\rm \textbf{q}\nu} = && \frac{1}{N(\epsilon_F)\omega_{\rm \textbf{q}\nu}} \sum_{nm} \int_{1^{st} BZ} \frac{d\rm \boldsymbol{k}}{\Omega_{BZ}} |g_{mn,\nu}(\rm \boldsymbol{k},\rm \boldsymbol{q})|^2  \nonumber\\
            & &\times \delta(\epsilon_{n\rm \boldsymbol{k}}-\epsilon_F)\delta(\epsilon_{m\rm \boldsymbol{k+q}}-\epsilon_F),
\end{eqnarray}

\noindent where $\epsilon$, $N(\epsilon_F)$, $\omega_{\rm \textbf{q}\nu}$ and $\Omega_{BZ}$ are the energy, DOS at Fermi energy, phonon frequency and the BZ volume, respectively. The above integration is done with the Gaussian smearing of 0.04 Ry for Dirac deltas\cite{nr20a}. Using the $\lambda_{\rm \textbf{q}\nu}$, we calculate the isotropic Eliashberg spectral function by taking the average electron-phonon interaction between an initial electronic state in the Fermi surface and all other states in the Fermi surface over all $\rm \textbf{q}$ in Fermi surface\cite{nr21}, as follows: 

\begin{eqnarray}
\alpha^2F(\omega)=\frac{1}{2}\sum_\nu\int_{1^{st} BZ} \frac{d\rm \boldsymbol{q}}{\Omega_{BZ}} \omega_{\rm \boldsymbol{q}\nu} \lambda_{\rm \boldsymbol{q}\nu} \delta(\omega-\omega_{\rm \boldsymbol{q}\nu})
\end{eqnarray}

Then the total electron-phonon coupling strength for a specific $\omega$ value, $\lambda(\omega)$ can be obtained by

\begin{eqnarray}
\lambda(\omega)=\int_{0}^\omega\frac{\alpha^2F(\omega')}{\omega'}d\omega'
\end{eqnarray}

When the above integration is taken till the highest $\omega$ value, i.e., $\omega_{max}$, we obtain $\lambda$ which is subsequently used to obtain the $T_{c}$ from the McMillan-Allen-Dynes formulae\cite{nr22},

\begin{eqnarray}
T_c = \frac{\omega_{log}}{1.2}\rm exp\bigg[\frac{-1.04(1+\lambda)}{\lambda-\mu^*_c(1+0.62\lambda)}\bigg]
\end{eqnarray}

\noindent where, $\mu^*_c$ is the effective Coulomb strength parameter and $\omega_{log}$ is the logarithmic averaged of the phonon frequency.


First we obtain the relaxed structure of monolayer silicene with position vectors of silicon atoms at ($1/3,2/3,z$) and ($2/3,1/3,z'$) in the unit cell. Then we construct the bilayer silicene in AA stacking and intercalate Ca atoms in between two layers, in such a way that the Ca atoms sit on top of the silicon hexagons. We denote the interlayer distance as d. Now the Ca atoms are sandwiched in a trigonal manner in between two silicene layers. The top and side views of the structure are depicted in Fig.1(a) and (b). We choose the AA stacked bilayer silicene due to its higher stability as compared to AB stacking, owing to reduced asymmetric hopping between two sublattice points in the unit cell\cite{nr23}. To obtain the distance of the Ca atom from the two silicene layers, we then vary the d, keeping equal distance of the silicene layers from the intercalant Ca atom and found the minimum energy structure with $d$ = 4.6 (see Fig.1(c)). For these calculations, we keep the lattice vector along the vertical direction, i.e., $c$ = 20\AA, that ensures that the bilayer sandwich compound is isolated. We perform further geometric relaxation with this minimum energy structure. These results in relaxed lattice constant of 3.815\AA, an interlayer distance of d = 4.704\AA, and the buckling of silicene layer $\rm \Delta$ = 0.646\AA. Note that, the buckling increases substantially as compared to the monolayer silicene where the $\rm \Delta$ is 0.44\AA. Moreover, the Ca and Si distance reduces in bilayer sandwich compound as compared to the ABC stacked bulk CaSi$_2$, indicating enhanced interaction between the Ca and silicene layers. The overall Si$_2$CaSi$_2$ structure finally shows a $D_{3d}$ point group symmetry.

\begin{figure}[h]
\includegraphics[scale=0.5]{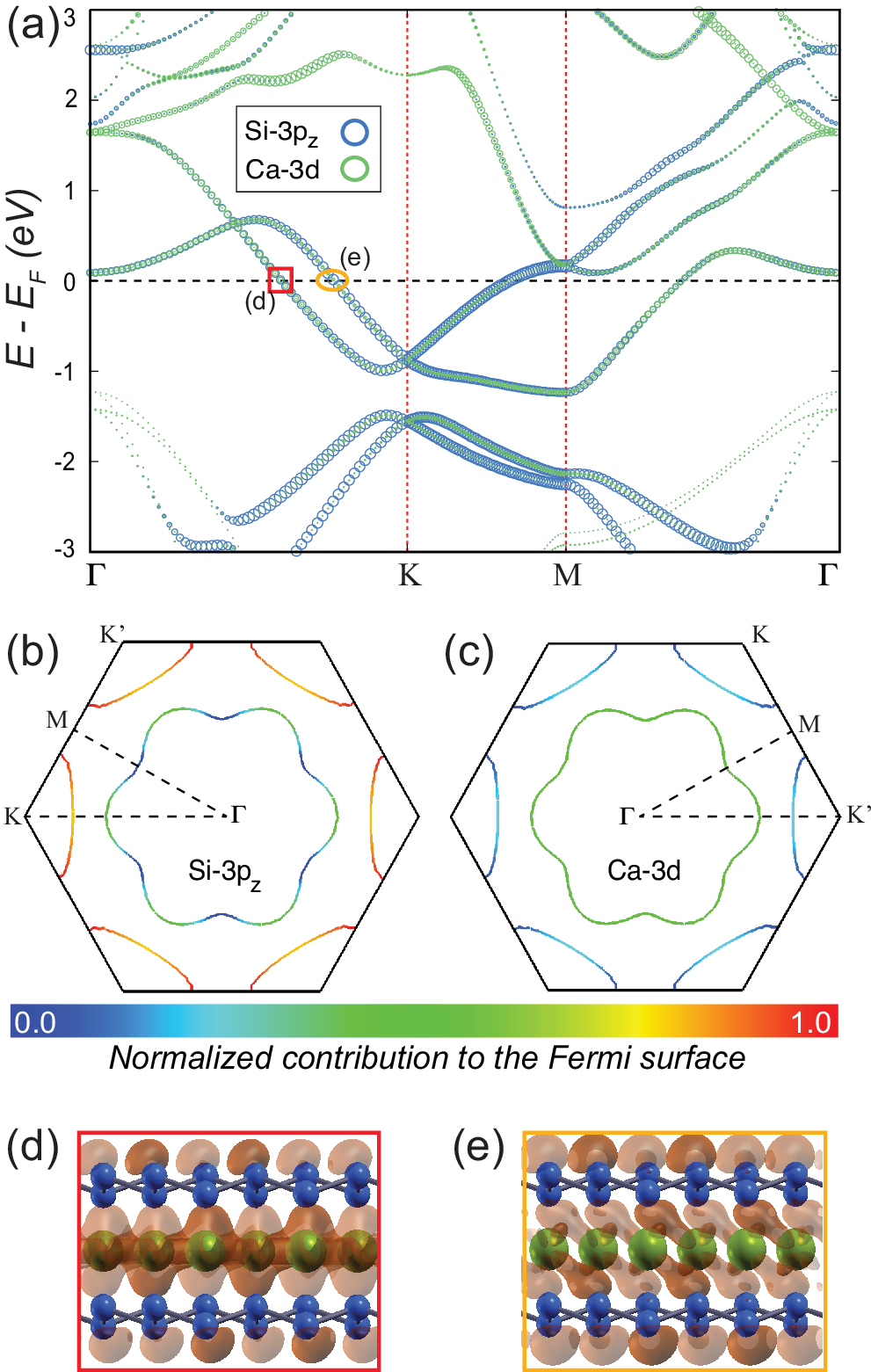}
\caption{\label{fig:epsart}  (color online) (a) The orbital projected scaled band structure of the Si$_2$CaSi$_2$ system, showing contributions from the Ca-3d and Si-3p$_{z}$ orbitals. The radius of the circles is proportional to the extent of contribution. The horizontal and vertical dashed lines depict the Fermi energy and the high-symmetric points, respectively. (b) and (c) present the contributions of Si-3p$_{z}$ and Ca-3d orbitals towards the Fermi surface in the hexagonal first Brillouin zone. (d) and (e) show the charge density distributions calculated at the Fermi energy for the two bands that cross the Fermi level between $\rm \Gamma$ and $\rm K$ points, as marked by a square and an oval in (a), respectively.}
\end{figure}

We calculate the electronic band structure of the Si$_2$CaSi$_2$ system and present the same in Fig.1(d), along with the DOS and orbital projected DOS. To get better insight of the role of Ca intercalant atoms, we also present the band structure of bilayer silicene in Si$_2$CaSi$_2$ geometry by removing the Ca atoms from the system (see dotted lines in Fig.1(d)). The Fermi levels of the above mentioned two systems are shown in terms of horizontal dashed and dashed-dotted lines. Note that, the Fermi level and the band structure of the bilayer silicene system are shifted with respect to the Fermi level of the Si$_2$CaSi$_2$ system for comparison. One can notice that the Fermi level the Si$_2$CaSi$_2$ system appears almost $\sim$1.5 eV above the Fermi level of the bilayer silicene system. This happens due to a partial charge transfer of 1.19$|e|$ from Ca to silicene layers, as revealed by the Bader charge analysis. Due to the charge transfer, the $\pi^*$ band near the high-symmetric point $K$ comes below the Fermi energy and creates an electron pocket.

It can be seen that, the intercalation of Ca substantially changes the band dispersions over the full irreducible Brillouin zone, indicating a considerably strong interlayer Ca-Si hybridization. The projected DOS shows large contributions from the Ca-3d and the Si-3p orbitals near the Fermi energy, whereas, the contributions from 4s and 4p orbitals of Ca have negligible contribution. This shows that the original empty d-orbitals of Ca is getting hybridized with the Si-3p orbitals and subsequently gets occupied. Increased Ca-3d occupancy also indicates a strong Ca-Si interaction\cite{nr24}. Note that, unlike the superconducting GIC systems, we do not observe any parabolic band crossing the Fermi energy near the $\rm \Gamma$ point. Such occupied parabolic interlayer band has been attributed to the superconducting property of the GIC systems\cite{nr25}. However, we observe such parabolic dispersion just above the Fermi level near the $\rm \Gamma$ point.

\begin{figure}[h]
\includegraphics[scale=0.38]{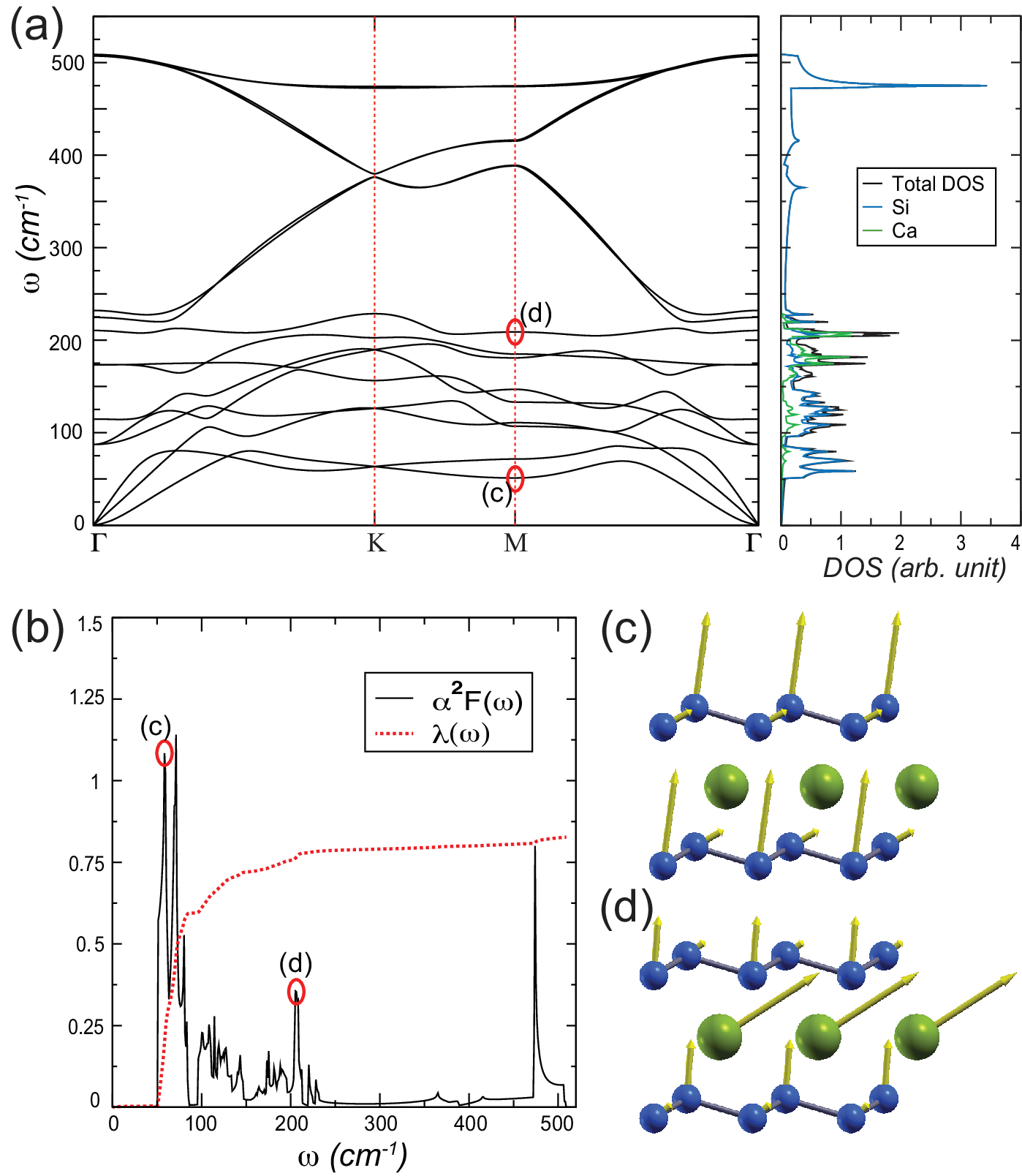}
\caption{\label{fig:epsart} (color online) (a) The phonon dispersion and the phonon DOS of the Si$_2$CaSi$_2$ system along with the contributions of Si and Ca orbitals therein. The vertical dashed lines show the location of the high-symmetric points. (b) The Eliashberg spectral function (solid line) along with the electron-phonon coupling strength, $\lambda (\omega)$ (dotted line) as a function of phonon frequency. (c) and (d) depict the directions of forces acting on each atom and the consequent out-of-plane vibration modes at two frequencies that contribute majorly towards the superconductivity, as marked in (a) and (b).}
\end{figure}

To gain further insight regarding the orbital contributions, we plot the Ca-3d and Si-3p$_z$ orbitals projected band structure in Fig.2(a). As can be seen, the parabolic band near the $\rm \Gamma$ point just above the Fermi energy shows equal contributions from both Ca-3d and Si-3p$_z$ orbitals. Therefore, this band is indeed the interlayer band that is known to induce superconductivity when gets occupied. However, in present system, this band remains empty and unlikely to contribute towards superconductivity in Si$_2$CaSi$_2$ system. Under experimental conditions, such band can get partially occupied and can give rise to small circular Fermi surface around the $\rm \Gamma$ point, as has been observed in case of GIC systems.

There are two other dispersive bands cross the Fermi level between $K$ and $\rm \Gamma$ points. One of the bands, marked with a square shows comparable contributions from Ca-3d and Si-3p$_z$ orbitals, revealing the hybridized nature of this band. The other band, marked with an oval exhibits major contributions from Si-3p$_z$ orbital. These two bands change their nature in other crystallographic directions. The first one exhibits predominantly Si-3p$_z$ contribution while crossing the Fermi level between $K$ and $M$ points, whereas the later one shows contributions from only Ca-3d orbital while crossing the Fermi level between $M$ and $\rm \Gamma$ points. 

In Fig.2(b) and (c), we present the normalized contributions of Si-3p$_z$ and Ca-3d towards the Fermi surface of the Si$_2$CaSi$_2$ system. As can be seen, the triangular electron pockets are created around the hexagonal Brillouin zone corners, i.e., around the $K$ and $K'$ points by Si-3p$_z$ orbitals. This arises due to the charge transfer to the silicene $\pi^*$ band from the Ca atoms. However, the flower like Fermi surface around the $\rm \Gamma$ point shows contributions from both the Si-3p$_z$ and Ca-3d orbitals, indicating its interlayer characteristics. For further insights, we plot the wave functions in Fig.2(d) and (e), calculated at the points marked with the square and oval in the band structure in Fig.2(a), respectively. The wave function for the interlayer band shows considerable overlap between the silicene layers and the Ca intercalant. However, the wave function for the occupied $\pi^*$ state shows localization over Si atoms with negligible overlap with the Ca atoms. The above analysis reveals the existence of the triangular electron pockets around the Brillouin zone corners, analogous to the GIC compounds. But instead of circular Fermi surface around the $\rm \Gamma$ point in GIC systems, we observe a flower like interlayer Fermi surface, arising from the hybridization between the Si $\pi^*$ states and the Ca-3d orbitals. Existence of such interlayer state indicates the possibility of superconductivity in Si$_2$CaSi$_2$ system.

To explore the superconducting properties, we investigate further the phonon behavior and subsequent electron-phonon coupling. We first present the phonon dispersion and the phonon DOS in Fig.3(a). The parabolic dispersion of the ZA acoustic mode near the $\rm \Gamma$ point is characteristic of the 2D system. The phonon DOS clearly indicates that the low energy phonon modes arise due to Si, which are actually out-of-plane vibrations. The Si in-plane vibrations give rise to the high energy optical modes. The phonon modes in bulk CaSi$_2$ are reported to get soften as compared to the monolayer silicene due to the charge transfer from the Ca to silicene layers\cite{nr26}. In bilayer sandwich compound Si$_2$CaSi$_2$, the phonon modes get further softened. At $K$ and $M$ points, the low energy acoustic phonon modes in Si$_2$CaSi$_2$ are 50-100 cm$^{-1}$ less than that of the bulk CaSi$_2$ and monolayer silicene\cite{nr27}. The E$_{2g}$ optical phonon modes at $\rm \Gamma$ point also get softened. The softening of the phonon modes are expected to enhance the electron-phonon coupling and thereby are expected to induce superconductivity in Si$_2$CaSi$_2$. On the other hand, the Ca vibrations actually contribute towards the low energy acoustic modes with first significant contribution starting around 180 cm$^{-1}$. The two major closely spaced peaks around 180 cm$^{-1}$ arise from the in-plane and out-of-plane vibrations of Ca intercalants. The next peak at 207 cm$^{-1}$, arising from the B$_{u}$ acoustic mode, shows considerable contributions from both Si and Ca out-of-plane vibrations. Therefore, we have three distinct regions phonon dispersions: (i) the low energy acoustic modes arising from the Si atoms, (ii) the higher energy acoustic modes arising from the Ca and Si coupling and (ii) high energy optical modes arising from the Si atoms.

In Fig.3(b), we plot the Eliashberg spectral function, $\alpha^2$F$(\omega)$, which provides the scattering amplitude of the electronic states on Fermi surface by a specific phonon mode (see Equation.2). The Eliashberg spectral function shows major contribution from the low-energy phonon modes that are arising from the out-of-plane Si vibrations. The maximum value of the $\lambda_{\boldsymbol{q}\nu}$ for the first peak at 50 cm$^{-1}$ is obtained from the B$_{u}$ acoustic mode at \textbf{M} point. We therefore present the vibration mode at this point in Fig.3(c). The force vectors clearly show the out-of-plane Si vibrations from both the silicene layers. The second peak at 70 cm$^{-1}$ also arises from the out-of-plane Si vibrations from both the silicene layers. The further higher frequency acoustic modes are also observed to contribute to the $\alpha^2$F$(\omega)$, with a large peak at 207 cm$^{-1}$. The highest value of $\lambda_{\boldsymbol{q}\nu}$ for this peak comes from the acoustic mode at \textbf{M} point. The force vectors at this point show out-of-plane vibrations from both the silicene layers and the Ca intercalant. Note that, the force vectors on Ca are inclined from the vertical direction, indicating the participation of both the in-plane and out-of-plane Ca vibrations.

We present the $\lambda(\omega)$ in Fig.3(b), that provides the cumulative electron-phonon coupling strength till the phonon frequency $\omega$ (see Equation.3). The $\lambda(\omega)$ also shows a rapid increase in the lower frequency regions, indicating strong electron-phonon coupling in this region, followed by saturation at higher frequency. A large peak of Eliashberg spectral function at 470 cm$^{-1}$ shows negligible contribution towards the $\lambda(\omega)$. Hence, this in-plane Si vibration mode in highest optical branch does not contribute to the electron-phonon coupling. Note that, the E$_{2g}$ mode at the $\rm \Gamma$ point and the A$_{1}$ mode at the \textbf{K} point for this optical branch does not exhibit Kohn anomaly, unlike in case of monolayer silicene, although in both cases this branch appears from the in-plane Si vibrations\cite{nr27}. The diverging phonon DOS corresponding to this branch appears due to flat phonon dispersion. However, the above mentioned modes exhibit larger phonon line width and a large peak in the Eliashberg spectral function due to the coupling with the Si $\pi^*$ electron density at Fermi energy. The negligible contribution of this peak towards the $\lambda(\omega)$ indicates the lack of participation of the Si $\pi^*$ state and the corresponding triangular Fermi pocket around the \textbf{K} towards the electron-phonon coupling.

For Si$_2$CaSi$_2$ system, we obtain the electron-phonon coupling strength, $\lambda$ = 0.85. Incorporating this value along with $\mu^*_c$ = 0.12\cite{nr8} in the McMillan-Allen-Dynes formulae (Equation.4)\cite{nr22}, we obtain the $T_{c}$ to be 5.4 K. Therefore, the superconductivity in Si$_2$CaSi$_2$ system arises from the coupling of the out-of-plane acoustic phonon modes of pure Si and combined Si and Ca with that of the interlayer flower like Fermi surface around the $\rm \Gamma$ point (Fig.2(b) and (c)). The softening of those acoustic modes due to the charge transfer from intercalant Ca to the silicene enhances such electron-phonon coupling, although the Si $\pi^*$ band does not contribute towards the superconductivity. {Consideration of higher spin-orbit coupling in this noncentrosymmetric system can further enhance the $T_{c}$, owing to the formation of energetically favorable non-$\rm \Gamma$-centric Cooper pairs\cite{nr32}}.


We investigate the superconducting property of Ca intercalated AA stacked bilayer silicene compound with the stoichiometry Si$_2$CaSi$_2$, using first principle calculations. Formation of a flower like interlayer Fermi surface around the $\rm \Gamma$ point enhances the density of states at the Fermi energy. Coupling of lower and intermediate frequency out-of-plane vibrations with such higher electron density near the Fermi energy, as evident from the Eliashberg spectral function, gives rise to superconductivity below 5.4 K. The $T_{c}$ is comparable with that of experimentally reported Li-decorated monolayer graphene\cite{nr28} and higher than the Ca intercalated bilayer graphene systems\cite{nr6}. The observation of superconductivity in Si$_2$CaSi$_2$ places it in the category of covalent superconductors, joining compounds like MgB$_2$\cite{nr29}, doped diamond\cite{nr30}, doped Si\cite{nr31}, and GIC\cite{nr4}, which maintain strong covalent bonds in their metallic state. Note that, the superconductivity in GIC systems has been explained in terms of the circular interlayer Fermi surface near the $\rm \Gamma$ point. In the present case, although the corresponding parabolic interlayer band remains unoccupied, the superconductivity arises due to higher hybridization of Si-3p$_z$ and Ca-3d orbitals and resulting dispersive bands away from the $\rm \Gamma$ point. Occupation of this band under pressure might result in increased $T_{c}$\cite{nr33}. These observations will make new inroads in achieving superconductivity in silicene based thin materials that can be easily integrated with the existing silicon technology for advanced applications.

\begin{acknowledgments}
JS thanks Ministry of Education, Govt. of India for PMRF grant (No. 901571). KW acknowledges the Japan Society for the Promotion of Science (JSPS) KAKENHI (Nos. 22H05473, JP21H01019, JP18H01154) and JST CREST (No. JPMJCR19T1) research grants. SD acknowledges IISER Tirupati for Intramural Funding and SERB, Department of Science and Technology (DST), Govt. of India for research grant CRG/2021/001731. JS and SD acknowledge National Supercomputing Mission (NSM) for providing computing resources of ‘PARAM Brahma’ at IISER Pune, which is implemented by C-DAC and supported by the Ministry of Electronics and Information Technology (MeitY) and DST, Govt. of India.
\end{acknowledgments}

\nocite{*}


\begin{thebibliography}{36}%
\makeatletter
\providecommand \@ifxundefined [1]{%
 \@ifx{#1\undefined}
}%
\providecommand \@ifnum [1]{%
 \ifnum #1\expandafter \@firstoftwo
 \else \expandafter \@secondoftwo
 \fi
}%
\providecommand \@ifx [1]{%
 \ifx #1\expandafter \@firstoftwo
 \else \expandafter \@secondoftwo
 \fi
}%
\providecommand \natexlab [1]{#1}%
\providecommand \enquote  [1]{``#1''}%
\providecommand \bibnamefont  [1]{#1}%
\providecommand \bibfnamefont [1]{#1}%
\providecommand \citenamefont [1]{#1}%
\providecommand \href@noop [0]{\@secondoftwo}%
\providecommand \href [0]{\begingroup \@sanitize@url \@href}%
\providecommand \@href[1]{\@@startlink{#1}\@@href}%
\providecommand \@@href[1]{\endgroup#1\@@endlink}%
\providecommand \@sanitize@url [0]{\catcode `\\12\catcode `\$12\catcode
  `\&12\catcode `\#12\catcode `\^12\catcode `\_12\catcode `\%12\relax}%
\providecommand \@@startlink[1]{}%
\providecommand \@@endlink[0]{}%
\providecommand \url  [0]{\begingroup\@sanitize@url \@url }%
\providecommand \@url [1]{\endgroup\@href {#1}{\urlprefix }}%
\providecommand \urlprefix  [0]{URL }%
\providecommand \Eprint [0]{\href }%
\providecommand \doibase [0]{https://doi.org/}%
\providecommand \selectlanguage [0]{\@gobble}%
\providecommand \bibinfo  [0]{\@secondoftwo}%
\providecommand \bibfield  [0]{\@secondoftwo}%
\providecommand \translation [1]{[#1]}%
\providecommand \BibitemOpen [0]{}%
\providecommand \bibitemStop [0]{}%
\providecommand \bibitemNoStop [0]{.\EOS\space}%
\providecommand \EOS [0]{\spacefactor3000\relax}%
\providecommand \BibitemShut  [1]{\csname bibitem#1\endcsname}%
\let\auto@bib@innerbib\@empty
\bibitem [{\citenamefont {Hirsch}\ and\ \citenamefont {Scalapino}(1986)}]{nr1}%
  \BibitemOpen
  \bibfield  {author} {\bibinfo {author} {\bibfnamefont {J.~E.}\ \bibnamefont
  {Hirsch}}\ and\ \bibinfo {author} {\bibfnamefont {D.~J.}\ \bibnamefont
  {Scalapino}},\ }\href@noop {} {\bibfield  {journal} {\bibinfo  {journal}
  {Phys. Rev. Lett.}\ }\textbf {\bibinfo {volume} {56}},\ \bibinfo {pages}
  {2732} (\bibinfo {year} {1986})}\BibitemShut {NoStop}%
\bibitem [{\citenamefont {Saito}\ \emph {et~al.}(2016)\citenamefont {Saito},
  \citenamefont {Nojima},\ and\ \citenamefont {Iwasa}}]{nr2}%
  \BibitemOpen
  \bibfield  {author} {\bibinfo {author} {\bibfnamefont {Y.}~\bibnamefont
  {Saito}}, \bibinfo {author} {\bibfnamefont {T.}~\bibnamefont {Nojima}},\ and\
  \bibinfo {author} {\bibfnamefont {Y.}~\bibnamefont {Iwasa}},\ }\bibfield
  {title} {\bibinfo {title} {Highly crystalline 2d superconductors},\
  }\href@noop {} {\bibfield  {journal} {\bibinfo  {journal} {Nat Rev Mater}\
  }\textbf {\bibinfo {volume} {2}},\ \bibinfo {pages} {16094} (\bibinfo {year}
  {2016})}\BibitemShut {NoStop}%
\bibitem [{\citenamefont {Qiu}\ \emph {et~al.}(2021)\citenamefont {Qiu},
  \citenamefont {Gong}, \citenamefont {Wang}, \citenamefont {Zhang},
  \citenamefont {Yang}, \citenamefont {Wang},\ and\ \citenamefont
  {Xiong}}]{nr3}%
  \BibitemOpen
  \bibfield  {author} {\bibinfo {author} {\bibfnamefont {D.}~\bibnamefont
  {Qiu}}, \bibinfo {author} {\bibfnamefont {C.}~\bibnamefont {Gong}}, \bibinfo
  {author} {\bibfnamefont {S.}~\bibnamefont {Wang}}, \bibinfo {author}
  {\bibfnamefont {M.}~\bibnamefont {Zhang}}, \bibinfo {author} {\bibfnamefont
  {C.}~\bibnamefont {Yang}}, \bibinfo {author} {\bibfnamefont {X.}~\bibnamefont
  {Wang}},\ and\ \bibinfo {author} {\bibfnamefont {J.}~\bibnamefont {Xiong}},\
  }\href@noop {} {\bibfield  {journal} {\bibinfo  {journal} {Adv. Mater.}\
  }\textbf {\bibinfo {volume} {33}},\ \bibinfo {pages} {2006124} (\bibinfo
  {year} {2021})}\BibitemShut {NoStop}%
\bibitem [{\citenamefont {Smith}\ \emph {et~al.}(2015)\citenamefont {Smith},
  \citenamefont {Weller}, \citenamefont {Howard}, \citenamefont {Dean},
  \citenamefont {Rahnejat}, \citenamefont {Saxena},\ and\ \citenamefont
  {Ellerby}}]{nr4}%
  \BibitemOpen
  \bibfield  {author} {\bibinfo {author} {\bibfnamefont {R.~P.}\ \bibnamefont
  {Smith}}, \bibinfo {author} {\bibfnamefont {T.~E.}\ \bibnamefont {Weller}},
  \bibinfo {author} {\bibfnamefont {C.~A.}\ \bibnamefont {Howard}}, \bibinfo
  {author} {\bibfnamefont {M.~P.}\ \bibnamefont {Dean}}, \bibinfo {author}
  {\bibfnamefont {K.~C.}\ \bibnamefont {Rahnejat}}, \bibinfo {author}
  {\bibfnamefont {S.~S.}\ \bibnamefont {Saxena}},\ and\ \bibinfo {author}
  {\bibfnamefont {M.}~\bibnamefont {Ellerby}},\ }\href@noop {} {\bibfield
  {journal} {\bibinfo  {journal} {Physica C}\ }\textbf {\bibinfo {volume}
  {514}},\ \bibinfo {pages} {50} (\bibinfo {year} {2015})}\BibitemShut
  {NoStop}%
\bibitem [{\citenamefont {Weller}\ \emph {et~al.}(2005)\citenamefont {Weller},
  \citenamefont {Ellerby}, \citenamefont {Saxena}, \citenamefont {Smith},\ and\
  \citenamefont {Skipper}}]{nr5}%
  \BibitemOpen
  \bibfield  {author} {\bibinfo {author} {\bibfnamefont {T.~E.}\ \bibnamefont
  {Weller}}, \bibinfo {author} {\bibfnamefont {M.}~\bibnamefont {Ellerby}},
  \bibinfo {author} {\bibfnamefont {S.~S.}\ \bibnamefont {Saxena}}, \bibinfo
  {author} {\bibfnamefont {R.~P.}\ \bibnamefont {Smith}},\ and\ \bibinfo
  {author} {\bibfnamefont {N.~T.}\ \bibnamefont {Skipper}},\ }\href@noop {}
  {\bibfield  {journal} {\bibinfo  {journal} {Nature Phys.}\ }\textbf {\bibinfo
  {volume} {1}},\ \bibinfo {pages} {39} (\bibinfo {year} {2005})}\BibitemShut
  {NoStop}%
\bibitem [{\citenamefont {Ichinokura}\ \emph {et~al.}(2016)\citenamefont
  {Ichinokura}, \citenamefont {Sugawara}, \citenamefont {Takayama},
  \citenamefont {Takahashi},\ and\ \citenamefont {Hasegawa}}]{nr6}%
  \BibitemOpen
  \bibfield  {author} {\bibinfo {author} {\bibfnamefont {S.}~\bibnamefont
  {Ichinokura}}, \bibinfo {author} {\bibfnamefont {K.}~\bibnamefont
  {Sugawara}}, \bibinfo {author} {\bibfnamefont {A.}~\bibnamefont {Takayama}},
  \bibinfo {author} {\bibfnamefont {T.}~\bibnamefont {Takahashi}},\ and\
  \bibinfo {author} {\bibfnamefont {S.}~\bibnamefont {Hasegawa}},\ }\href@noop
  {} {\bibfield  {journal} {\bibinfo  {journal} {ACS Nano}\ }\textbf {\bibinfo
  {volume} {10}},\ \bibinfo {pages} {2761} (\bibinfo {year}
  {2016})}\BibitemShut {NoStop}%
\bibitem [{\citenamefont {Zheng}\ and\ \citenamefont {Margine}(2016)}]{nr7}%
  \BibitemOpen
  \bibfield  {author} {\bibinfo {author} {\bibfnamefont {J.-J.}\ \bibnamefont
  {Zheng}}\ and\ \bibinfo {author} {\bibfnamefont {E.~R.}\ \bibnamefont
  {Margine}},\ }\href@noop {} {\bibfield  {journal} {\bibinfo  {journal} {Phys.
  Rev. B}\ }\textbf {\bibinfo {volume} {94}},\ \bibinfo {pages} {064509}
  (\bibinfo {year} {2016})}\BibitemShut {NoStop}%
\bibitem [{\citenamefont {Calandra}\ and\ \citenamefont {Mauri}(2005)}]{nr8}%
  \BibitemOpen
  \bibfield  {author} {\bibinfo {author} {\bibfnamefont {M.}~\bibnamefont
  {Calandra}}\ and\ \bibinfo {author} {\bibfnamefont {F.}~\bibnamefont
  {Mauri}},\ }\href@noop {} {\bibfield  {journal} {\bibinfo  {journal} {Phys.
  Rev. Lett.}\ }\textbf {\bibinfo {volume} {95}},\ \bibinfo {pages} {237002}
  (\bibinfo {year} {2005})}\BibitemShut {NoStop}%
\bibitem [{\citenamefont {Csányi}\ \emph {et~al.}(2005)\citenamefont
  {Csányi}, \citenamefont {Littlewood}, \citenamefont {Nevidomskyy},
  \citenamefont {Pickard},\ and\ \citenamefont {Simons}}]{nr9}%
  \BibitemOpen
  \bibfield  {author} {\bibinfo {author} {\bibfnamefont {G.}~\bibnamefont
  {Csányi}}, \bibinfo {author} {\bibfnamefont {P.~B.}\ \bibnamefont
  {Littlewood}}, \bibinfo {author} {\bibfnamefont {A.~H.}\ \bibnamefont
  {Nevidomskyy}}, \bibinfo {author} {\bibfnamefont {C.~J.}\ \bibnamefont
  {Pickard}},\ and\ \bibinfo {author} {\bibfnamefont {B.~D.}\ \bibnamefont
  {Simons}},\ }\href@noop {} {\bibfield  {journal} {\bibinfo  {journal} {Nature
  Phys.}\ }\textbf {\bibinfo {volume} {1}},\ \bibinfo {pages} {42} (\bibinfo
  {year} {2005})}\BibitemShut {NoStop}%
\bibitem [{\citenamefont {Vogt}\ \emph {et~al.}(2012)\citenamefont {Vogt},
  \citenamefont {De~Padova}, \citenamefont {Quaresima}, \citenamefont {Avila},
  \citenamefont {Frantzeskakis}, \citenamefont {Asensio}, \citenamefont
  {Resta}, \citenamefont {Ealet},\ and\ \citenamefont {Le~Lay}}]{nr9a}%
  \BibitemOpen
  \bibfield  {author} {\bibinfo {author} {\bibfnamefont {P.}~\bibnamefont
  {Vogt}}, \bibinfo {author} {\bibfnamefont {P.}~\bibnamefont {De~Padova}},
  \bibinfo {author} {\bibfnamefont {C.}~\bibnamefont {Quaresima}}, \bibinfo
  {author} {\bibfnamefont {J.}~\bibnamefont {Avila}}, \bibinfo {author}
  {\bibfnamefont {E.}~\bibnamefont {Frantzeskakis}}, \bibinfo {author}
  {\bibfnamefont {M.~C.}\ \bibnamefont {Asensio}}, \bibinfo {author}
  {\bibfnamefont {A.}~\bibnamefont {Resta}}, \bibinfo {author} {\bibfnamefont
  {B.}~\bibnamefont {Ealet}},\ and\ \bibinfo {author} {\bibfnamefont
  {G.}~\bibnamefont {Le~Lay}},\ }\href@noop {} {\bibfield  {journal} {\bibinfo
  {journal} {Phys. Rev. Lett.}\ }\textbf {\bibinfo {volume} {108}},\ \bibinfo
  {pages} {155501} (\bibinfo {year} {2012})}\BibitemShut {NoStop}%
\bibitem [{\citenamefont {Takeda}\ and\ \citenamefont
  {Shiraishi}(1994)}]{nr9b}%
  \BibitemOpen
  \bibfield  {author} {\bibinfo {author} {\bibfnamefont {K.}~\bibnamefont
  {Takeda}}\ and\ \bibinfo {author} {\bibfnamefont {K.}~\bibnamefont
  {Shiraishi}},\ }\href@noop {} {\bibfield  {journal} {\bibinfo  {journal}
  {Phys. Rev. B}\ }\textbf {\bibinfo {volume} {50}},\ \bibinfo {pages} {14916}
  (\bibinfo {year} {1994})}\BibitemShut {NoStop}%
\bibitem [{\citenamefont {Fischetti}\ and\ \citenamefont
  {Vandenberghe}(2016)}]{nr10}%
  \BibitemOpen
  \bibfield  {author} {\bibinfo {author} {\bibfnamefont {M.~V.}\ \bibnamefont
  {Fischetti}}\ and\ \bibinfo {author} {\bibfnamefont {W.~G.}\ \bibnamefont
  {Vandenberghe}},\ }\href@noop {} {\bibfield  {journal} {\bibinfo  {journal}
  {Phys. Rev. B}\ }\textbf {\bibinfo {volume} {93}},\ \bibinfo {pages} {155413}
  (\bibinfo {year} {2016})}\BibitemShut {NoStop}%
\bibitem [{\citenamefont {Ma\~nes}(2007)}]{nr11}%
  \BibitemOpen
  \bibfield  {author} {\bibinfo {author} {\bibfnamefont {J.~L.}\ \bibnamefont
  {Ma\~nes}},\ }\href@noop {} {\bibfield  {journal} {\bibinfo  {journal} {Phys.
  Rev. B}\ }\textbf {\bibinfo {volume} {76}},\ \bibinfo {pages} {045430}
  (\bibinfo {year} {2007})}\BibitemShut {NoStop}%
\bibitem [{\citenamefont {Jishi}\ \emph {et~al.}(1991)\citenamefont {Jishi},
  \citenamefont {Dresselhaus},\ and\ \citenamefont {Chaiken}}]{nr12}%
  \BibitemOpen
  \bibfield  {author} {\bibinfo {author} {\bibfnamefont {R.~A.}\ \bibnamefont
  {Jishi}}, \bibinfo {author} {\bibfnamefont {M.~S.}\ \bibnamefont
  {Dresselhaus}},\ and\ \bibinfo {author} {\bibfnamefont {A.}~\bibnamefont
  {Chaiken}},\ }\href@noop {} {\bibfield  {journal} {\bibinfo  {journal} {Phys.
  Rev. B}\ }\textbf {\bibinfo {volume} {44}},\ \bibinfo {pages} {10248}
  (\bibinfo {year} {1991})}\BibitemShut {NoStop}%
\bibitem [{\citenamefont {Wan}\ \emph {et~al.}(2013)\citenamefont {Wan},
  \citenamefont {Ge}, \citenamefont {Yang},\ and\ \citenamefont {Yao}}]{nr13}%
  \BibitemOpen
  \bibfield  {author} {\bibinfo {author} {\bibfnamefont {W.}~\bibnamefont
  {Wan}}, \bibinfo {author} {\bibfnamefont {Y.}~\bibnamefont {Ge}}, \bibinfo
  {author} {\bibfnamefont {F.}~\bibnamefont {Yang}},\ and\ \bibinfo {author}
  {\bibfnamefont {Y.}~\bibnamefont {Yao}},\ }\href@noop {} {\bibfield
  {journal} {\bibinfo  {journal} {EPL}\ }\textbf {\bibinfo {volume} {104}},\
  \bibinfo {pages} {36001} (\bibinfo {year} {2013})}\BibitemShut {NoStop}%
\bibitem [{\citenamefont {Chen}\ \emph {et~al.}(2013)\citenamefont {Chen},
  \citenamefont {Feng},\ and\ \citenamefont {Wu}}]{nr14}%
  \BibitemOpen
  \bibfield  {author} {\bibinfo {author} {\bibfnamefont {L.}~\bibnamefont
  {Chen}}, \bibinfo {author} {\bibfnamefont {B.}~\bibnamefont {Feng}},\ and\
  \bibinfo {author} {\bibfnamefont {K.}~\bibnamefont {Wu}},\ }\href@noop {}
  {\bibfield  {journal} {\bibinfo  {journal} {Appl. Phys. Lett.}\ }\textbf
  {\bibinfo {volume} {102}},\ \bibinfo {pages} {081602} (\bibinfo {year}
  {2013})}\BibitemShut {NoStop}%
\bibitem [{\citenamefont {Feng}\ \emph {et~al.}(2012)\citenamefont {Feng},
  \citenamefont {Ding}, \citenamefont {Meng}, \citenamefont {Yao},
  \citenamefont {He}, \citenamefont {Cheng}, \citenamefont {Chen},\ and\
  \citenamefont {Wu}}]{nr15}%
  \BibitemOpen
  \bibfield  {author} {\bibinfo {author} {\bibfnamefont {B.}~\bibnamefont
  {Feng}}, \bibinfo {author} {\bibfnamefont {Z.}~\bibnamefont {Ding}}, \bibinfo
  {author} {\bibfnamefont {S.}~\bibnamefont {Meng}}, \bibinfo {author}
  {\bibfnamefont {Y.}~\bibnamefont {Yao}}, \bibinfo {author} {\bibfnamefont
  {X.}~\bibnamefont {He}}, \bibinfo {author} {\bibfnamefont {P.}~\bibnamefont
  {Cheng}}, \bibinfo {author} {\bibfnamefont {L.}~\bibnamefont {Chen}},\ and\
  \bibinfo {author} {\bibfnamefont {K.}~\bibnamefont {Wu}},\ }\href@noop {}
  {\bibfield  {journal} {\bibinfo  {journal} {Nano Letters}\ }\textbf {\bibinfo
  {volume} {12}},\ \bibinfo {pages} {3507} (\bibinfo {year}
  {2012})}\BibitemShut {NoStop}%
\bibitem [{\citenamefont {Affronte}\ \emph {et~al.}(1998)\citenamefont
  {Affronte}, \citenamefont {Laborde}, \citenamefont {Olcese},\ and\
  \citenamefont {Palenzona}}]{nr16}%
  \BibitemOpen
  \bibfield  {author} {\bibinfo {author} {\bibfnamefont {M.}~\bibnamefont
  {Affronte}}, \bibinfo {author} {\bibfnamefont {O.}~\bibnamefont {Laborde}},
  \bibinfo {author} {\bibfnamefont {G.}~\bibnamefont {Olcese}},\ and\ \bibinfo
  {author} {\bibfnamefont {A.}~\bibnamefont {Palenzona}},\ }\href@noop {}
  {\bibfield  {journal} {\bibinfo  {journal} {Journal of Alloys and Compounds}\
  }\textbf {\bibinfo {volume} {274}},\ \bibinfo {pages} {68} (\bibinfo {year}
  {1998})}\BibitemShut {NoStop}%
\bibitem [{\citenamefont {Sanfilippo}\ \emph {et~al.}(2000)\citenamefont
  {Sanfilippo}, \citenamefont {Elsinger}, \citenamefont {N\'u\~nez-Regueiro},
  \citenamefont {Laborde}, \citenamefont {LeFloch}, \citenamefont {Affronte},
  \citenamefont {Olcese},\ and\ \citenamefont {Palenzona}}]{nr17}%
  \BibitemOpen
  \bibfield  {author} {\bibinfo {author} {\bibfnamefont {S.}~\bibnamefont
  {Sanfilippo}}, \bibinfo {author} {\bibfnamefont {H.}~\bibnamefont
  {Elsinger}}, \bibinfo {author} {\bibfnamefont {M.}~\bibnamefont {N\'u\~nez
  Regueiro}}, \bibinfo {author} {\bibfnamefont {O.}~\bibnamefont {Laborde}},
  \bibinfo {author} {\bibfnamefont {S.}~\bibnamefont {LeFloch}}, \bibinfo
  {author} {\bibfnamefont {M.}~\bibnamefont {Affronte}}, \bibinfo {author}
  {\bibfnamefont {G.~L.}\ \bibnamefont {Olcese}},\ and\ \bibinfo {author}
  {\bibfnamefont {A.}~\bibnamefont {Palenzona}},\ }\href@noop {} {\bibfield
  {journal} {\bibinfo  {journal} {Phys. Rev. B}\ }\textbf {\bibinfo {volume}
  {61}},\ \bibinfo {pages} {R3800} (\bibinfo {year} {2000})}\BibitemShut
  {NoStop}%
\bibitem [{\citenamefont {Giannozzi}\ \emph {et~al.}(2009)\citenamefont
  {Giannozzi}, \citenamefont {Baroni}, \citenamefont {Bonini}, \citenamefont
  {Calandra}, \citenamefont {Car}, \citenamefont {Cavazzoni}, \citenamefont
  {Ceresoli}, \citenamefont {Chiarotti}, \citenamefont {Cococcioni},
  \citenamefont {Dabo}, \citenamefont {Corso}, \citenamefont {de~Gironcoli},
  \citenamefont {Fabris}, \citenamefont {Fratesi}, \citenamefont {Gebauer},
  \citenamefont {Gerstmann}, \citenamefont {Gougoussis}, \citenamefont
  {Kokalj}, \citenamefont {Lazzeri}, \citenamefont {Martin-Samos},
  \citenamefont {Marzari}, \citenamefont {Mauri}, \citenamefont {Mazzarello},
  \citenamefont {Paolini}, \citenamefont {Pasquarello}, \citenamefont
  {Paulatto}, \citenamefont {Sbraccia}, \citenamefont {Scandolo}, \citenamefont
  {Sclauzero}, \citenamefont {Seitsonen}, \citenamefont {Smogunov},
  \citenamefont {Umari},\ and\ \citenamefont {Wentzcovitch}}]{nr18}%
  \BibitemOpen
  \bibfield  {author} {\bibinfo {author} {\bibfnamefont {P.}~\bibnamefont
  {Giannozzi}}, \bibinfo {author} {\bibfnamefont {S.}~\bibnamefont {Baroni}},
  \bibinfo {author} {\bibfnamefont {N.}~\bibnamefont {Bonini}}, \bibinfo
  {author} {\bibfnamefont {M.}~\bibnamefont {Calandra}}, \bibinfo {author}
  {\bibfnamefont {R.}~\bibnamefont {Car}}, \bibinfo {author} {\bibfnamefont
  {C.}~\bibnamefont {Cavazzoni}}, \bibinfo {author} {\bibfnamefont
  {D.}~\bibnamefont {Ceresoli}}, \bibinfo {author} {\bibfnamefont {G.~L.}\
  \bibnamefont {Chiarotti}}, \bibinfo {author} {\bibfnamefont {M.}~\bibnamefont
  {Cococcioni}}, \bibinfo {author} {\bibfnamefont {I.}~\bibnamefont {Dabo}},
  \bibinfo {author} {\bibfnamefont {A.~D.}\ \bibnamefont {Corso}}, \bibinfo
  {author} {\bibfnamefont {S.}~\bibnamefont {de~Gironcoli}}, \bibinfo {author}
  {\bibfnamefont {S.}~\bibnamefont {Fabris}}, \bibinfo {author} {\bibfnamefont
  {G.}~\bibnamefont {Fratesi}}, \bibinfo {author} {\bibfnamefont
  {R.}~\bibnamefont {Gebauer}}, \bibinfo {author} {\bibfnamefont
  {U.}~\bibnamefont {Gerstmann}}, \bibinfo {author} {\bibfnamefont
  {C.}~\bibnamefont {Gougoussis}}, \bibinfo {author} {\bibfnamefont
  {A.}~\bibnamefont {Kokalj}}, \bibinfo {author} {\bibfnamefont
  {M.}~\bibnamefont {Lazzeri}}, \bibinfo {author} {\bibfnamefont
  {L.}~\bibnamefont {Martin-Samos}}, \bibinfo {author} {\bibfnamefont
  {N.}~\bibnamefont {Marzari}}, \bibinfo {author} {\bibfnamefont
  {F.}~\bibnamefont {Mauri}}, \bibinfo {author} {\bibfnamefont
  {R.}~\bibnamefont {Mazzarello}}, \bibinfo {author} {\bibfnamefont
  {S.}~\bibnamefont {Paolini}}, \bibinfo {author} {\bibfnamefont
  {A.}~\bibnamefont {Pasquarello}}, \bibinfo {author} {\bibfnamefont
  {L.}~\bibnamefont {Paulatto}}, \bibinfo {author} {\bibfnamefont
  {C.}~\bibnamefont {Sbraccia}}, \bibinfo {author} {\bibfnamefont
  {S.}~\bibnamefont {Scandolo}}, \bibinfo {author} {\bibfnamefont
  {G.}~\bibnamefont {Sclauzero}}, \bibinfo {author} {\bibfnamefont {A.~P.}\
  \bibnamefont {Seitsonen}}, \bibinfo {author} {\bibfnamefont {A.}~\bibnamefont
  {Smogunov}}, \bibinfo {author} {\bibfnamefont {P.}~\bibnamefont {Umari}},\
  and\ \bibinfo {author} {\bibfnamefont {R.~M.}\ \bibnamefont {Wentzcovitch}},\
  }\bibfield  {title} {\bibinfo {title} {Quantum espresso: a modular and
  open-source software project for quantum simulations of materials},\
  }\href@noop {} {\bibfield  {journal} {\bibinfo  {journal} {J. Phys.: Condens.
  Matter}\ }\textbf {\bibinfo {volume} {21}},\ \bibinfo {pages} {395502}
  (\bibinfo {year} {2009})}\BibitemShut {NoStop}%
\bibitem [{\citenamefont {Ceperley}\ and\ \citenamefont {Alder}(1980)}]{nr19}%
  \BibitemOpen
  \bibfield  {author} {\bibinfo {author} {\bibfnamefont {D.~M.}\ \bibnamefont
  {Ceperley}}\ and\ \bibinfo {author} {\bibfnamefont {B.~J.}\ \bibnamefont
  {Alder}},\ }\href@noop {} {\bibfield  {journal} {\bibinfo  {journal} {Phys.
  Rev. Lett.}\ }\textbf {\bibinfo {volume} {45}},\ \bibinfo {pages} {566}
  (\bibinfo {year} {1980})}\BibitemShut {NoStop}%
\bibitem [{\citenamefont {Hartwigsen}\ \emph {et~al.}(1998)\citenamefont
  {Hartwigsen}, \citenamefont {Goedecker},\ and\ \citenamefont
  {Hutter}}]{nr20}%
  \BibitemOpen
  \bibfield  {author} {\bibinfo {author} {\bibfnamefont {C.}~\bibnamefont
  {Hartwigsen}}, \bibinfo {author} {\bibfnamefont {S.}~\bibnamefont
  {Goedecker}},\ and\ \bibinfo {author} {\bibfnamefont {J.}~\bibnamefont
  {Hutter}},\ }\href@noop {} {\bibfield  {journal} {\bibinfo  {journal} {Phys.
  Rev. B}\ }\textbf {\bibinfo {volume} {58}},\ \bibinfo {pages} {3641}
  (\bibinfo {year} {1998})}\BibitemShut {NoStop}%
\bibitem [{\citenamefont {Bl\"ochl}(1994)}]{nr20a}%
  \BibitemOpen
  \bibfield  {author} {\bibinfo {author} {\bibfnamefont {P.~E.}\ \bibnamefont
  {Bl\"ochl}},\ }\href@noop {} {\bibfield  {journal} {\bibinfo  {journal}
  {Phys. Rev. B}\ }\textbf {\bibinfo {volume} {50}},\ \bibinfo {pages} {17953}
  (\bibinfo {year} {1994})}\BibitemShut {NoStop}%
\bibitem [{\citenamefont {Allen}\ and\ \citenamefont {Mitrović}(1983)}]{nr21}%
  \BibitemOpen
  \bibfield  {author} {\bibinfo {author} {\bibfnamefont {P.~B.}\ \bibnamefont
  {Allen}}\ and\ \bibinfo {author} {\bibfnamefont {B.}~\bibnamefont
  {Mitrović}},\ }\bibfield  {title} {\bibinfo {title} {Theory of
  superconducting tc}\ }(\bibinfo  {publisher} {Academic Press},\ \bibinfo
  {year} {1983})\ pp.\ \bibinfo {pages} {1--92}\BibitemShut {NoStop}%
\bibitem [{\citenamefont {McMillan}(1968)}]{nr22}%
  \BibitemOpen
  \bibfield  {author} {\bibinfo {author} {\bibfnamefont {W.~L.}\ \bibnamefont
  {McMillan}},\ }\href@noop {} {\bibfield  {journal} {\bibinfo  {journal}
  {Phys. Rev.}\ }\textbf {\bibinfo {volume} {167}},\ \bibinfo {pages} {331}
  (\bibinfo {year} {1968})}\BibitemShut {NoStop}%
\bibitem [{\citenamefont {Dutta}\ and\ \citenamefont
  {Wakabayashi}(2015)}]{nr23}%
  \BibitemOpen
  \bibfield  {author} {\bibinfo {author} {\bibfnamefont {S.}~\bibnamefont
  {Dutta}}\ and\ \bibinfo {author} {\bibfnamefont {K.}~\bibnamefont
  {Wakabayashi}},\ }\href@noop {} {\bibfield  {journal} {\bibinfo  {journal}
  {Phys. Rev. B}\ }\textbf {\bibinfo {volume} {91}},\ \bibinfo {pages} {201410(R)}
  (\bibinfo {year} {2015})}\BibitemShut {NoStop}%
\bibitem [{\citenamefont {Kulatov}\ \emph {et~al.}(1997)\citenamefont
  {Kulatov}, \citenamefont {Nakayama},\ and\ \citenamefont {Ohta}}]{nr24}%
  \BibitemOpen
  \bibfield  {author} {\bibinfo {author} {\bibfnamefont {E.}~\bibnamefont
  {Kulatov}}, \bibinfo {author} {\bibfnamefont {H.}~\bibnamefont {Nakayama}},\
  and\ \bibinfo {author} {\bibfnamefont {H.}~\bibnamefont {Ohta}},\ }\href@noop
  {} {\bibfield  {journal} {\bibinfo  {journal} {J. Phys.: Condens. Matter}\
  }\textbf {\bibinfo {volume} {9}},\ \bibinfo {pages} {10159} (\bibinfo {year}
  {1997})}\BibitemShut {NoStop}%
\bibitem [{\citenamefont {Sugawara}\ \emph {et~al.}(2009)\citenamefont
  {Sugawara}, \citenamefont {Sato},\ and\ \citenamefont {Takahashi}}]{nr25}%
  \BibitemOpen
  \bibfield  {author} {\bibinfo {author} {\bibfnamefont {K.}~\bibnamefont
  {Sugawara}}, \bibinfo {author} {\bibfnamefont {T.}~\bibnamefont {Sato}},\
  and\ \bibinfo {author} {\bibfnamefont {T.}~\bibnamefont {Takahashi}},\
  }\href@noop {} {\bibfield  {journal} {\bibinfo  {journal} {Nature Phys}\
  }\textbf {\bibinfo {volume} {5}},\ \bibinfo {pages} {1745} (\bibinfo {year}
  {2009})}\BibitemShut {NoStop}%
\bibitem [{\citenamefont {Cheng}\ \emph {et~al.}(2011)\citenamefont {Cheng},
  \citenamefont {Zhu},\ and\ \citenamefont {Schwingenschlögl}}]{nr26}%
  \BibitemOpen
  \bibfield  {author} {\bibinfo {author} {\bibfnamefont {Y.~C.}\ \bibnamefont
  {Cheng}}, \bibinfo {author} {\bibfnamefont {Z.~Y.}\ \bibnamefont {Zhu}},\
  and\ \bibinfo {author} {\bibfnamefont {U.}~\bibnamefont
  {Schwingenschlögl}},\ }\href@noop {} {\bibfield  {journal} {\bibinfo
  {journal} {EPL}\ }\textbf {\bibinfo {volume} {95}},\ \bibinfo {pages} {17005}
  (\bibinfo {year} {2011})}\BibitemShut {NoStop}%
\bibitem [{\citenamefont {Yan}\ \emph {et~al.}(2013)\citenamefont {Yan},
  \citenamefont {Stein}, \citenamefont {Schaefer}, \citenamefont {Wang},\ and\
  \citenamefont {Chou}}]{nr27}%
  \BibitemOpen
  \bibfield  {author} {\bibinfo {author} {\bibfnamefont {J.-A.}\ \bibnamefont
  {Yan}}, \bibinfo {author} {\bibfnamefont {R.}~\bibnamefont {Stein}}, \bibinfo
  {author} {\bibfnamefont {D.~M.}\ \bibnamefont {Schaefer}}, \bibinfo {author}
  {\bibfnamefont {X.-Q.}\ \bibnamefont {Wang}},\ and\ \bibinfo {author}
  {\bibfnamefont {M.~Y.}\ \bibnamefont {Chou}},\ }\href@noop {} {\bibfield
  {journal} {\bibinfo  {journal} {Phys. Rev. B}\ }\textbf {\bibinfo {volume}
  {88}},\ \bibinfo {pages} {121403(R)} (\bibinfo {year} {2013})}\BibitemShut
  {NoStop}%
\bibitem [{\citenamefont {Smidman}\ \emph {et~al.}(2017)\citenamefont
  {Smidman}, \citenamefont {Salamon}, \citenamefont {Yuan},\ and\ \citenamefont
  {Agterberg}}]{nr32}%
  \BibitemOpen
  \bibfield  {author} {\bibinfo {author} {\bibfnamefont {M.}~\bibnamefont
  {Smidman}}, \bibinfo {author} {\bibfnamefont {M.~B.}\ \bibnamefont
  {Salamon}}, \bibinfo {author} {\bibfnamefont {H.~Q.}\ \bibnamefont {Yuan}},\
  and\ \bibinfo {author} {\bibfnamefont {D.~F.}\ \bibnamefont {Agterberg}},\
  }\href@noop {} {\bibfield  {journal} {\bibinfo  {journal} {Reports on
  Progress in Physics}\ }\textbf {\bibinfo {volume} {80}},\ \bibinfo {pages}
  {036501} (\bibinfo {year} {2017})}\BibitemShut {NoStop}%
\bibitem [{\citenamefont {Ludbrook}\ \emph {et~al.}(2015)\citenamefont
  {Ludbrook}, \citenamefont {Levy}, \citenamefont {Nigge}, \citenamefont
  {Zonno}, \citenamefont {Schneider}, \citenamefont {Dvorak}, \citenamefont
  {Veenstra}, \citenamefont {Zhdanovich}, \citenamefont {Wong}, \citenamefont
  {Dosanjh}, \citenamefont {Straßer}, \citenamefont {Stöhr}, \citenamefont
  {Forti}, \citenamefont {Ast}, \citenamefont {Starke},\ and\ \citenamefont
  {Damascelli}}]{nr28}%
  \BibitemOpen
  \bibfield  {author} {\bibinfo {author} {\bibfnamefont {B.~M.}\ \bibnamefont
  {Ludbrook}}, \bibinfo {author} {\bibfnamefont {G.}~\bibnamefont {Levy}},
  \bibinfo {author} {\bibfnamefont {P.}~\bibnamefont {Nigge}}, \bibinfo
  {author} {\bibfnamefont {M.}~\bibnamefont {Zonno}}, \bibinfo {author}
  {\bibfnamefont {M.}~\bibnamefont {Schneider}}, \bibinfo {author}
  {\bibfnamefont {D.~J.}\ \bibnamefont {Dvorak}}, \bibinfo {author}
  {\bibfnamefont {C.~N.}\ \bibnamefont {Veenstra}}, \bibinfo {author}
  {\bibfnamefont {S.}~\bibnamefont {Zhdanovich}}, \bibinfo {author}
  {\bibfnamefont {D.}~\bibnamefont {Wong}}, \bibinfo {author} {\bibfnamefont
  {P.}~\bibnamefont {Dosanjh}}, \bibinfo {author} {\bibfnamefont
  {C.}~\bibnamefont {Straßer}}, \bibinfo {author} {\bibfnamefont
  {A.}~\bibnamefont {Stöhr}}, \bibinfo {author} {\bibfnamefont
  {S.}~\bibnamefont {Forti}}, \bibinfo {author} {\bibfnamefont {C.~R.}\
  \bibnamefont {Ast}}, \bibinfo {author} {\bibfnamefont {U.}~\bibnamefont
  {Starke}},\ and\ \bibinfo {author} {\bibfnamefont {A.}~\bibnamefont
  {Damascelli}},\ }\href@noop {} {\bibfield  {journal} {\bibinfo  {journal}
  {Proceedings of the National Academy of Sciences}\ }\textbf {\bibinfo
  {volume} {112}},\ \bibinfo {pages} {11795} (\bibinfo {year}
  {2015})}\BibitemShut {NoStop}%
\bibitem [{\citenamefont {Nagamatsu}\ \emph {et~al.}(2004)\citenamefont
  {Nagamatsu}, \citenamefont {Nakagawa}, \citenamefont {Muranaka},
  \citenamefont {Zenitani},\ and\ \citenamefont {Akimitsu}}]{nr29}%
  \BibitemOpen
  \bibfield  {author} {\bibinfo {author} {\bibfnamefont {J.}~\bibnamefont
  {Nagamatsu}}, \bibinfo {author} {\bibfnamefont {N.}~\bibnamefont {Nakagawa}},
  \bibinfo {author} {\bibfnamefont {T.}~\bibnamefont {Muranaka}}, \bibinfo
  {author} {\bibfnamefont {Y.}~\bibnamefont {Zenitani}},\ and\ \bibinfo
  {author} {\bibfnamefont {J.}~\bibnamefont {Akimitsu}},\ }\href@noop {}
  {\bibfield  {journal} {\bibinfo  {journal} {Nature}\ }\textbf {\bibinfo
  {volume} {410}},\ \bibinfo {pages} {63} (\bibinfo {year} {2004})}\BibitemShut
  {NoStop}%
\bibitem [{\citenamefont {Ekimov}\ \emph {et~al.}(2004)\citenamefont {Ekimov},
  \citenamefont {Sidorov}, \citenamefont {Bauer}, \citenamefont {Mel'nik},
  \citenamefont {Curro}, \citenamefont {Thompson},\ and\ \citenamefont
  {Stishov}}]{nr30}%
  \BibitemOpen
  \bibfield  {author} {\bibinfo {author} {\bibfnamefont {E.~A.}\ \bibnamefont
  {Ekimov}}, \bibinfo {author} {\bibfnamefont {V.~A.}\ \bibnamefont {Sidorov}},
  \bibinfo {author} {\bibfnamefont {E.~D.}\ \bibnamefont {Bauer}}, \bibinfo
  {author} {\bibfnamefont {N.~N.}\ \bibnamefont {Mel'nik}}, \bibinfo {author}
  {\bibfnamefont {N.~J.}\ \bibnamefont {Curro}}, \bibinfo {author}
  {\bibfnamefont {J.~D.}\ \bibnamefont {Thompson}},\ and\ \bibinfo {author}
  {\bibfnamefont {S.~M.}\ \bibnamefont {Stishov}},\ }\href@noop {} {\bibfield
  {journal} {\bibinfo  {journal} {Nature}\ }\textbf {\bibinfo {volume} {428}},\
  \bibinfo {pages} {542} (\bibinfo {year} {2004})}\BibitemShut {NoStop}%
\bibitem [{\citenamefont {Bustarret}\ \emph {et~al.}(2006)\citenamefont
  {Bustarret}, \citenamefont {Marcenat}, \citenamefont {Achatz}, \citenamefont
  {Kačmarčik}, \citenamefont {Lévy}, \citenamefont {Huxley}, \citenamefont
  {Ortéga}, \citenamefont {Bourgeois}, \citenamefont {Blase}, \citenamefont
  {Débarre},\ and\ \citenamefont {Boulmer}}]{nr31}%
  \BibitemOpen
  \bibfield  {author} {\bibinfo {author} {\bibfnamefont {E.}~\bibnamefont
  {Bustarret}}, \bibinfo {author} {\bibfnamefont {C.}~\bibnamefont {Marcenat}},
  \bibinfo {author} {\bibfnamefont {P.}~\bibnamefont {Achatz}}, \bibinfo
  {author} {\bibfnamefont {J.}~\bibnamefont {Kačmarčik}}, \bibinfo {author}
  {\bibfnamefont {F.}~\bibnamefont {Lévy}}, \bibinfo {author} {\bibfnamefont
  {A.}~\bibnamefont {Huxley}}, \bibinfo {author} {\bibfnamefont
  {L.}~\bibnamefont {Ortéga}}, \bibinfo {author} {\bibfnamefont
  {E.}~\bibnamefont {Bourgeois}}, \bibinfo {author} {\bibfnamefont
  {X.}~\bibnamefont {Blase}}, \bibinfo {author} {\bibfnamefont
  {D.}~\bibnamefont {Débarre}},\ and\ \bibinfo {author} {\bibfnamefont
  {J.}~\bibnamefont {Boulmer}},\ }\href@noop {} {\bibfield  {journal} {\bibinfo
   {journal} {Nature}\ }\textbf {\bibinfo {volume} {444}},\ \bibinfo {pages}
  {465} (\bibinfo {year} {2006})}\BibitemShut {NoStop}%
\bibitem [{\citenamefont {Belash}\ \emph {et~al.}(1989)\citenamefont {Belash},
  \citenamefont {Bronnikov}, \citenamefont {Zharikov},\ and\ \citenamefont
  {Pal'nichenko}}]{nr33}%
  \BibitemOpen
  \bibfield  {author} {\bibinfo {author} {\bibfnamefont {I.}~\bibnamefont
  {Belash}}, \bibinfo {author} {\bibfnamefont {A.}~\bibnamefont {Bronnikov}},
  \bibinfo {author} {\bibfnamefont {O.}~\bibnamefont {Zharikov}},\ and\
  \bibinfo {author} {\bibfnamefont {A.}~\bibnamefont {Pal'nichenko}},\
  }\href@noop {} {\bibfield  {journal} {\bibinfo  {journal} {Solid State
  Communications}\ }\textbf {\bibinfo {volume} {69}},\ \bibinfo {pages} {921}
  (\bibinfo {year} {1989})}\BibitemShut {NoStop}%
\end{thebibliography}

%

\end{document}